\documentstyle[prl,aps]{revtex}
\begin{document}
\twocolumn[
%\draft
\preprint{U. of MD PP \#95--110}
\preprint{DOE/ER/40762-057}
\title {\bf Double Penguins and the Contribution of
Vector Meson--like States to the Decays
$B \rightarrow K^* \gamma, \, B \rightarrow \rho \gamma$}
\author{J. Milana}
\address{Department of Physics, University of Maryland,
College Park, Maryland 20742, USA}
\date{DOE/ER/40762-057, U. of MD PP \#95--110,
 March, 1995, revised October, 1995}
\maketitle
\vskip -0.2in
\begin{abstract} \widetext
Using perturbative QCD,  the contribution at the leading twist,
leading  $\alpha_s$ level, of charm and up quark loops
to the decays $B \rightarrow K^* \gamma$ and $B \rightarrow \rho \gamma$
is presented.   In the case of $B \rightarrow \rho \gamma$, the relative
importance
of these contributions depend upon the unknown CKM matrix elements
$V_{bu}$ and $V_{td}$.    Assuming that the  ratio
$r = V_{bc}V^*_{cd}/V_{bt}V^*_{td}$ is bounded between
$-2.25 \le r \le -.5$ as is suggested by the Particle Data Group,
the error in extracting $ |V_{td}/V_{ts}| $ by these decays is estimated.
\end{abstract}
\vskip 0.2in
%\pacs{PACS numbers: 13.20.Hw, 12.38.Bx, 14.40.Nd}
]
%\newpage
\narrowtext
\section{Introduction}
The recent observation\cite{cleo93} of the rare decay $B \rightarrow K^*
\gamma$
was the first unambiguous experimental verification of flavor--changing
neutral currents.
While not occurring at tree--level in the Standard Model, such
currents, or ``penguins'',
 are well--known\cite{fcnc} to arise due to loop effects.  The magnitude
of these transitions depend\cite{ILim} upon fundamental parameters in the
Standard Model such as the mass of the top quark as well as the CKM matrix
elements $V_{tb} V_{ts}^*$, although significant QCD enhancements\cite{rgroup}
tend to offset somewhat the dependence on the former.
A systematic determination of the two--body rare decays of the $B$ meson,
both for determining these fundamental parameters
as well as for indications of possible physics beyond the Standard
Model, forms a major
part of the goals of the future $B$ factories at SLAC and KEK.
Despite these connections, significant theoretical input
will nevertheless be necessary to interpret the data so as to
distinguish purely hadronic effects as well as potentially competing
mechanisms.

With the latest data on $B \rightarrow J/\psi \, K^*$\cite{charmdata},
the question of the relative importance\cite{VMD1} of
``long--distance'' effects arising from intermediate charmonium states to the
decay
$B \rightarrow K^* \gamma$ has received new attention\cite{VMD2}.
The potential relevance of annihilation diagrams in the decay
$B \rightarrow \rho \gamma$ has also been emphasized by
Atwood {\it et al}.\cite{VMD2}
as a possible source of contamination of the extraction using these two decay
modes\cite{Alisugg} of the ratio $|V_{td}/V_{ts}|$.  While these effects are
undoubtedly present, it is crucial for their analysis that all mechanisms are
understood within a single formalism.  For example, most of the studies
cited employ
vector--meson dominance to estimate from the measured charmonium decay rate a
 contribution to the $B \rightarrow K^* \gamma$ transition amplitude
purportedly
distinct from the $b \rightarrow s \gamma$ penguin vertex already present
in the effective Hamiltonian.   However the (significant) running of the Wilson
coefficient of the electromagnetic penguin operator in the effective
Hamiltonian
is predominantly driven by its mixing with other operators via charm--quark
loops.
There is hence a serious potential of double--counting when using a mixed
meson--quark language to describe the transition amplitude.  This
potential is further
emphasized when recalling that the mixing between $O_7$ and $O_2$
(see Eqs.~(\ref{empengop}) and (\ref{4ptop}) below) is in fact identically
zero at the one--loop level.\cite{rgroup,AliGreub}

Recently, perturbative QCD (pQCD) methods were applied to
the penguin decay $B \rightarrow K^* \gamma$\cite{cmpeng}.  Previously pQCD had
 been found\cite{cmsystem} to be quite successful\cite{shb} in describing
the hadronic, two--body decay channels of the $B$ meson.
The large mass scale of the decaying
$B$ meson, coupled with the restriction to the two body exclusive decay modes
involving nearly massless, highly Lorentz contracted states,
implies that the transition amplitude is governed by short--distance processes
(on hadronic scales) and hence an appropriate environment to apply pQCD.
Indeed such simple kinematic considerations raises the question as to
the true suitability of calling any mechanism contributing
to these decays ``long--distance'' and further emphasizes the need for
a unified approach.

The dominant graph in the pQCD framework that contributes to the decay
$B \rightarrow K^* \gamma$ is shown in Fig.~(1).
A branching ratio of roughly  $3 \times 10^{-5}$ was obtained in
Ref.~\cite{cmpeng},
in reasonable agreement with the data\cite{cleo93}.   One process that was
however omitted in that analysis was the so--called double penguin
graphs of Fig.~(2).
Based on the analysis of Ali and Greub\cite{AliGreub} to the
decay $b \rightarrow s \, \gamma \, g$, it was argued that such
graphs would be suppressed compared
to the dominant decay mechanism shown in Fig.~(1).  The fact that the
virtuality of
the gluon in Fig.~(2) is in general significantly off--shell (of
$O(m_c^2)$ in practice)
and is embedded in additional loops,
implies that the analytical expressions found in Ref. \cite{AliGreub}
using on--shell gluon emission could not be simply applied.
A reasonable foundation for omitting the graph was therefore welcomed
and exploited.

For a detailed understanding of the decay amplitude, and especially for the
extrapolation to the case $B \rightarrow \rho \gamma$ where CKM factors no
longer
suppress other background processes, it is nevertheless important
to quantify these additional mechanisms.
This paper reports the results of such an analysis.
While as we will see, these results do in fact support the conjecture that
these
mechanisms enhance the total decay rate, nontrivial hadronic phases
enter into obtaining
this result.  This possibility has been generally ignored.  The enhancement is
related to the observed\cite{charmdata} failure of factorization
in the decay $B \rightarrow J/\psi K^*$,
although the overall enhancement factor of the decay rate itself
depends sensitively
on details of the wavefunction of the $K^*$ in a fashion that the
dominant decay mechanism, Fig. (1), does not.

\section{Calculations}
Exclusive processes at large momentum transfer are
addressed \cite{BrodLep} within pQCD starting with a Fock component
expansion of the involved hadrons whereby a twist expansion suggests that
 the contribution from the lowest order Fock component
dominates the physical observable under consideration.
An exclusive process then involves a perturbatively calculable hard
amplitude convoluted with a nonperturbative, soft physics wavefunction,
$\psi_m$, from each of the hadrons $m$ entering or leaving the hard
interaction.  These wavefunctions,
although as yet uncalculable from first principles,
are universal for each meson, {\it i.e.}
they factorize from the hard amplitude and hence are independent of
the process involved.   In exclusive processes in pQCD
they play the analogous role that structure functions do in the case
of inclusive scattering events.
Thus as was employed in Ref. \cite{cmsystem}, ideally one can
phenomenologically
parametrize these wavefunctions using a (few) measured cross--sections/decay
rates.

The factorization scheme
advocated by Brodsky and Lepage \cite{BrodLep} is employed, whereby
 the momenta of the quarks are taken as some fraction $x$ of the
total momentum of the parent meson weighted by a soft physics distribution
amplitude $\phi(x)$.
The peaking approximation is used for $\phi_B$,
the distribution amplitude of the $B$ meson, wherein
\begin{equation}
\phi_B(x) = {f_B \over 2\sqrt 3} \delta (x-\epsilon_B).
\end{equation}
The decay constant of the $B$ is $f_B$ (in the convention $f_\pi = 93$MeV)
and $x$ is the light cone momentum fraction carried by the light quark.
The parameter $\epsilon_B$ in $\phi_B(x)$ is related to the difference
in the masses of the $B$ meson and $b$-quark,
\begin{equation}
m_B = m_b + \bar\Lambda_B
\label{massofB}
\end{equation}
whereby $\epsilon_B = \bar\Lambda_B / m_B$.

Ignoring the mass of the $K^*$, its
distribution amplitude can be written as
\begin{equation}
\phi_{K^*}(y) = {\sqrt 3} f_{K^*} y (1-y) \tilde \phi_{K^*}(y).
\end{equation}
Two guesses for $\phi_{K^*}(y)$ were considered in \cite{cmpeng},
\begin{eqnarray}
\phi_{K^*}(y) &=& 1\nonumber\\
\phi_{K^*}(y) &=&  5 \,  y^2 \, (1 - y)^2.
\label{distAmps}
\end{eqnarray}
The first is the so--called super-asymptotic\cite{asympt} distribution
amplitude,
while the lower form is one suggested by Chernyak, Zhitnitsky and
Zhitnitsky (CZZ)\cite{czz} for the transverse polarizations of the $\rho$
meson.
\footnote{See however Ref.~\cite{Alisugg} for a QCD sum rule result for
 $\phi_{K^*}(y)$}

In the present context, the factorization scheme is
augmented by the viability of an $\epsilon_B$ expansion for the decay
amplitude.
All terms in the hard amplitude of order $\epsilon_B^2$ are ignored, both
because they are expected to be numerically small and because they are related,
through the mass of the light quark, to transverse momentum effects.  Since the
latter is ignored in the factorization scheme, self--consistency dictates these
other terms also be ignored.  The parameter  $\epsilon_B$  has
been fitted\cite{cmsystem,cm93} using the decay $B \rightarrow D \pi$.  With
mild
assumptions concerning the decay constants $f_B$ and $f_D$, a typical value
found
was $\epsilon_B = .095$.

\subsection{Dominant mechanism}
In Ref. \cite{cmpeng}, Fig.~(1) was shown to dominate $all$ graphs
involving penguin
 operators.  To leading order in $\epsilon_B$, the contribution to the decay
$B^+ \rightarrow K^{*+} \gamma$  was found to be
\begin{eqnarray}
M_{\gamma{\rm peng}}&&=
-{8G V_{tb} V_{ts}^*\over m_B \epsilon_B} \alpha_s(\mu) C_7(\mu) I
\nonumber \\
   &&\times       \left( p \cdot q \epsilon^* \cdot \xi^*
          + i\epsilon_{\mu\nu\alpha\beta}
          p^{\mu} q^{\nu} \epsilon^{* \alpha} \xi^{* \beta} \right),
\label{leadingamp}
\end{eqnarray}
where $\epsilon$ ($\xi$) is the polarization of the photon ($K^*$),
\begin{equation}
G = {e C_F \over 4\pi} {G_F\over\sqrt 2} f_B f_{K^*},
\end{equation}
and $C_7(\mu)$ is the renormalization group improved\cite{rgroup}
Wilson coefficient of the electromagnetic penguin operator
\begin{equation}
O_7 = {e\over 16\pi^2}\ m_b \bar s \sigma^{\mu \nu} F_{\mu \nu}
{1\over 2} \left (1+\gamma_5 \right) b.
\label{empengop}
\end{equation}
The quantity $I$ involves an integral over the distribution
amplitude of the $K^*$ and is given by
\begin{equation}
I = \int_0^1 dy\,\tilde\phi_{K^*}(y)\,
         {(1-y)(1+y-2\epsilon_B)\over y-2\epsilon_B-i0^+}.
\end{equation}
For the two distribution functions considered, this becomes
(using $\epsilon_B =. 095$ as discussed earlier)
\begin{equation}
I = \left\{ \begin{array}{c}.68 + i \, 2.55\\1.75 + i \, 1.96\end{array}\right.
,
\label{pengint}
\end{equation}
where the upper number is for the asymptotic distribution and the
lower is for the CZZ one.  The imaginary part comes from an internal propagator
going on--shell, kinematically allowed here by the inputed
soft--physics parametrization
Eq.~(\ref{massofB})  that $m_b < M_B$.   As in other cases in
pQCD\cite{examples},
it is a calculable hadronic phase since the overall
kinematics of the reaction dictate that only short distance propagation occurs,
as discussed in~\cite{cm93}.  In more technical language, the pole is
not pinched and
hence not associated with a long distance event~\cite{cn65+}.

One curious result of Eq.~(\ref{pengint}) is that $I^2$ is nearly
identical despite the
significant differences in phase for
the two distribution amplitudes under consideration.  Hence the total decay
rate
 was found in \cite{cmpeng} insensitive to these soft physics inputs.
This pattern however does not continue when additional
mechanisms are included in which interference effects depend
crucially on the details of the relative phases.

\subsection{ The Annihilation graphs}
The first such competing mechanism, the annihilation diagrams of Fig.~(3)
 was already considered in \cite{cmpeng}.
\footnote{The importance of these annihilation diagrams  was first realized by
Bander {\it et al.} in Ref.~\cite{history} in the context of $D$ meson decays
using a nonrelativistic quark model approach.}
As suggested by the figure, the dominant result (in $1/\epsilon_B$) is from the
graph involving photon emission
off of the light--quark of the $B$ meson.   To this order in $\epsilon_B$, the
contribution of this process to the decay $B^+ \rightarrow K^{*+} \gamma$ is
\begin{eqnarray}
M_{\rm ann} &=&  -\frac{2e_u \left(C_2(\tilde\mu)+
\frac{1}{N_c}C_1(\tilde\mu) \right)}
{m_B \epsilon_B }  \,
 {m_{K^*} \over m_B} \nonumber\\
&&\quad\times \left[ e {G_F \over\sqrt 2}
V_{ub} V_{us}^* f_B f_{K^*}\right]\nonumber \\
&&\quad \times  \left( p \cdot q \epsilon^* \cdot \xi^*
+ i\epsilon_{\mu\nu\alpha\beta} p^{\mu} q^{\nu}
\epsilon^{* \alpha} \xi^{* \beta} \right), \label{anniheq}
\end{eqnarray}
where $e_u = 2/3$ and $C_2$ and $C_1$ are the Wilson coefficients of the
4--point operators
(greek subscripts are color indices)
\begin{eqnarray}
O_1 &=& \frac{1}{4}\bar u_\alpha \gamma^\mu (1-\gamma_5) b_\beta \,
\bar s_\beta \gamma_\mu (1-\gamma_5) u_\alpha,\nonumber\\
O_2 &=& \frac{1}{4}\bar u_\alpha \gamma^\mu (1-\gamma_5) b_\alpha \,
\bar s_\beta \gamma_\mu (1-\gamma_5) u_\beta.\label{4ptop}
\end{eqnarray}
In this case $m_{K^*}$ is kept as it appears as an
overall factor arising from the $W$ turning into a $K^*$.
Due to the differences in CKM matrix elements as well as the factor
$m_K^*/M_B$,
the annihilation amplitude is here essentially ignorable.
As though discussed by Cheng and also Atwood {\it et al.}\cite{VMD2}, this
is no longer true for the (as yet unseen) decay mode $B \rightarrow \rho \,
\gamma$.
For the decay $B^+ \rightarrow \rho^+ \, \gamma$, $M_{ann}$ is obtained
from Eq.~(\ref{anniheq}) with the obvious modification in CKM factors and
meson decay constant.   For  the neutral decay $B^0 \rightarrow \rho^0 \,
\gamma$,
\begin{eqnarray}
M_{\rm ann} &=&  -\frac{2e_d \left(C_1(\tilde\mu)+
\frac{1}{N_c}C_2(\tilde\mu) \right)}
{m_B \epsilon_B }  \,
 {m_{K^*} \over m_B} \nonumber\\
&&\quad\times\left[ e {G_F \over\sqrt 2}
V_{ub} V_{ud}^* f_B \frac{f_{\rho}}{\sqrt2}\right]\nonumber \\
&&\quad \times  \left( p \cdot q \epsilon^* \cdot \xi^*
+ i\epsilon_{\mu\nu\alpha\beta} p^{\mu} q^{\nu}
\epsilon^{* \alpha} \xi^{* \beta} \right).
\end{eqnarray}

\subsection{The Quark-Loop graphs}
We now come to the second competing mechanism, the
 quark--loop graphs of Fig. (2).    These have been recently considered
by Greub {\it et al.} in Ref.~\cite{CPstuff} using a quark model approach.
These authors were predominantly interested in studying CP violating effects
and thus focussed on only the absorptive parts of these graphs.  Here though,
the
entire amplitude is of interest.  The important application
to CP violation will be deferred to a later work.

In the evaluation of the graphs of Fig. (2), one finds that
while each graph is individually ultraviolet divergent, their sum is finite.
Likewise gauge--invariance (in both the strong and electromagnetic
interactions) is only obtained after the graphs are summed.
In the case that the gluon is on--shell, $Q^2 = 0$, entirely
analytical results\cite{AliGreub} are possible.
Such a convenient form has not been found in the present context which
involves a second
integration over the gluon's virtuality, $Q^2 = -y \epsilon_B M_B^2$.
Intermediate results will be presented, in which the remaining
integrals were then evaluated
numerically.  The contribution of the
two quark loop graphs of Fig. (2), $M_{q\bar q\,{\rm loop}}$, to the
decay amplitude is
\begin{eqnarray}
M_{q\bar q\,{\rm loop}}&= &-\frac{16 G V_{qb} V_{qs}^*}
{3 m_B \epsilon_B}\alpha_s(\tilde\mu)
	C_2(\tilde\mu) \int_0^1 dy\,\tilde\phi_{K^*}(y)\,\tilde I (y)\nonumber\\
&&\times \left( p \cdot q \epsilon^* \cdot \xi^* +
i\epsilon_{\mu\nu\alpha\beta}
          p^{\mu} q^{\nu} \epsilon^{* \alpha} \xi^{* \beta} \right),
\label{charmamp}
\end{eqnarray}
where $\tilde I (y)$ is
\begin{eqnarray}
\tilde I &(y)& = \frac{3}{4}
+ \frac{1}{2(Q^2)^2}\left(m_q^4 \ln\left|1+\frac{Q^2}{m_q^2}\right| -
Q^2 m_q^2\right)
\nonumber\\
&+& \int_0^1 dx\, \left( \frac{-m^2_q}{2q\cdot Q x}
\ln\left|\frac{m^2_q + Q^2 x (x-1)}{m^2_q + 2q\cdot Q x (x-1)}\right|\right.
\nonumber\\
&&\hspace{.5in}+ (1-x)\ln\left|\frac{m^2_q + 2q\cdot Q x
(x-1)}{m_q^2+Q^2}\right|
\nonumber\\
&&\hspace{.5in} + \left.\frac{2q\cdot Q x^2}{\beta}
\ln\left|\frac{(1-z_+)(x-z_-)}{(1-z_-)(x-z_+)}\right|\right)
\nonumber\\
&+&\frac{i\pi}{2} \Theta(1 + \frac{m^2_q}{Q^2})
\left[ 1 - \left( \frac{m^2_q}{Q^2}\right)^2 \right]
- i \pi \Theta(y-y_o)\times\nonumber\\
&\biggl{[} & \frac{\sqrt{1-4\alpha}}{2} +
\frac{m^2_q}{2q\cdot Q} \ln\frac{x^+}{x^-}
 - 2q\cdot Q \int_{x^-}^{x^+} dx\,
\frac{x^2}{\beta}\biggl{]},
\end{eqnarray}
 in which
\begin{eqnarray}
&y_o = \epsilon_B + \frac{4 m_q^2}{M_B^2},
\hspace{.3in}&\alpha = \frac{m^2_q}{M_B^2 [ y  - \epsilon_B ]},
\nonumber\\
&z_{\pm} = \frac{- 2q\cdot Q x \pm \beta}{2 Q^2},
\hspace{.3in}
&x^{\pm} = \frac{1 \pm \sqrt{1 - 4\alpha}}{2}
\end{eqnarray}
and $\beta = \sqrt{(2q\cdot Q x)^2 - 4 Q^2 m_q^2}$.
To obtain this result, it is important to strictly keep only terms that
are leading in $\epsilon_B^{-1}$.
Subleading terms are higher--twist and require other higher--twist elements
(such as transverse momentum degrees of freedom) to maintain gauge--invariance.

Numerical integration yields that
\begin{equation}
\int_0^1 dy\,\tilde\phi_{K^*}(y)\,\tilde I (y) =
\left\{ \begin{array}{cc}.51 + i \, .14\hspace{.1in}&.71 - i \, .30\\
.43 + i \, .61\hspace{.1in}&.64 - i \, .30\end{array}\right. ,
\label{qloopres}
\end{equation}
where the results are presented as in Eq. (\ref{pengint}).  The two columns are
the results for the case of the up ($m_q = 0$) and charm quark loops
($m_c = 1.5 GeV$)
respectively.   There is negligible change in the charm quark results taking
 $1.25 GeV< m_c < 1.75 GeV$.

\section{Results}
For the decay $B \rightarrow K^* \, \gamma$, the two relevant amplitudes are
$M_{\gamma{\rm peng}}$ and $M_{c\bar c\,{\rm loop}}$.
As indicated by their expressions, in order
to reflect the fact that the average virtuality of the exchanged gluon
($Q^2$) in each
of the mechanisms is different, the running coupling and Wilson coefficient are
evaluated at  scales appropriate to each amplitude.   For the dominant piece,
$M_{\gamma{\rm peng}}$, this occurs at $\mu^2 \approx .5$GeV$^2$; $-Q^2$ at the
pole in the bottom quark's propagator.   In the case of $M_{c\bar c\,{\rm
loop}}$,
$\langle -Q^2 \rangle \approx \frac{1}{2} \epsilon_B M_B^2 = 1.3$GeV$^2$.
Note that a change in scale $\mu^2$ affects the
overall magnitude of the branching rate predominantly by the dependence
on $\alpha_s$ in Eq.~(\ref{leadingamp}).

For $\Lambda_{QCD} = .2$GeV one obtains for the
enhancement factor $R$ in the total decay rate the result that
\begin{equation}
R = \frac{(M_{\gamma{\rm peng}} + M_{c\bar c\,{\rm loop}})^2}
{M_{\gamma{\rm peng}}^2} = \left\{ \begin{array}{c}1.06\\1.31\end{array}\right.
,
\end{equation}
where the two results are for the asymptotic and CZZ distribution amplitudes
respectively.    Note that in this ratio the uncertainty due to our
present ignorance
of $f_B$, as well as the dominant dependence on the parameter $\epsilon_B$
cancel.
There is little dependence in $R$ on the exact value of  $\Lambda_{QCD}$.
However a precise value for $R$ is clearly dependent upon details of the
kaon's soft--physics information (unlike the square modulus of Eq.
(\ref{pengint})).
The range is due primarily to the phase differences in $M_{\gamma{\rm peng}}$
(Eq. (\ref{qloopres}) shows that
$M_{c\bar c\,{\rm loop}}$ is nearly insensitive to the choice of
$\phi_{K^*}(y)$).
One should note that this range is comparable to the various estimates
obtained using vector--meson dominance methods\cite{VMD1,VMD2}.  Such
duality is perhaps best understood by further noting that in
pQCD\cite{cmnonfact}
the nonfactorizing amplitudes are found to produce large contributions to the
transversally polarized final states of the decay $B \rightarrow J/\psi
\, K^{(*)}$.

In the case $B^+ \rightarrow \rho^+ \, \gamma$, CKM factors no longer
suppress either
the annihilation diagrams or the up--quark loop contributions.    To
quantify this dependence,
unitarity of CKM matrix is exploited, whereby
\begin{equation}
V_{bu} V^*_{ud}+V_{bc} V^*_{cd} + V_{bt}V^*_{td} = 0.
\end{equation}
Defining the ratio $V_{bc}V^*_{cd}/V_{bt}V^*_{td} = r$ and assuming
SU(3) symmetric distribution amplitudes for the $\rho$ and $K^*$ mesons
one obtains for the ratio of decay rates that
\begin{equation}
\frac{\Gamma_{B^+ \rightarrow \rho^+ \, \gamma}}
{\Gamma_{B^+ \rightarrow K^{*+} \, \gamma}}
= \left(\frac{f_\rho V_{td}}{f_{K^*} V_{ts}}\right)^2  f_{asy,CZZ}(r),
\label{decrats}
\end{equation}
where $f_{asy,CZZ}(r) \ne 1$ represents the error due
to competing mechanisms of extracting this ratio.  The subscripts
 recall the fact that $f(r)$ is dependent upon the  distribution amplitude
used for the  $\rho$ and $K^*$ mesons.
In Fig. (4), $f(r=-2.25)$ and $f(r=-.5)$ are plotted as a function of
$\Lambda_{QCD}$
(the range of $r$ as suggested by the Particle Data Group is
$-2.25 \le r \le -.5$,\cite{partdata}).    We observe that while $f(r)$ is
fairly independent of the distribution amplitudes, there is significant
dependence on $\Lambda_{QCD}$.    Note that the deviation of $f(r)$ from $1$
is greatest for smallest $\Lambda_{QCD}$, where one would expect that the
perturbative formalism employed here is more accurate.
A conservative estimate is therefore that these these decay modes can be used
to obtain $| V_{td}/V_{ts} |$\cite{Alisugg} to only within a factor of two or
so.

In the case of the neutral $B$ decays, the effective absence of the
annihilation diagram
 improves upon this result.   From Fig. (5) we obtain an estimated uncertainty
in extracting $|V_{td}/V_{ts}|$ on the order of $50\%$.

\section{Conclusion}
The two body exclusive decays of the $B$ meson necessarily involves modeling of
the initial and final state hadrons.
Perturbative QCD methods have been previously shown\cite{cmsystem}
to be a robust framework for describing these processes and earlier
work\cite{cmpeng}
had already shown that from the dominant decay mechanism, Fig.~(1),
the observed rate for $B \rightarrow K^* \gamma$\cite{cleo93}
was consistent with a pQCD approach.   Using this experience,
the question of the relative importance of subdominant processes to the decays
$B \rightarrow K^* \gamma$ and $B \rightarrow \rho \gamma$ have been herein
addressed at the leading twist, leading $\alpha_s$ level.
By working within a single, coherent framework the possibility of double
counting present when using a mixed parton--meson approach has been avoided.

In the case of $B \rightarrow K^* \gamma$,
subdominant charm quark loops\cite{AliGreub},
$M_{c\bar c\,{\rm loop}}$, here called double--penguins
(and elsewhere in the literature\cite{VMD1,VMD2}
as ``long--distance'',  vector meson like states)
have been found to yield an enhancement of up to $30$\%
in the total decay rate, depending upon bound--state parameters of the $K^*$.
The uncertainty is due primarily to phase differences between
$M_{\gamma{\rm peng}}$,
the dominant mechanism, and $M_{c\bar c\,{\rm loop}}$.

In the case of $B \rightarrow \rho \gamma$, which depends upon the unknown CKM
matrix elements $V_{bu}$ and $V_{td}$, additional subdominant processes
involving up quark loops and also annihilation diagrams\cite{history}
may no longer be CKM
suppressed.      Assuming $SU_f(3)$ symmetry of the $\rho$ and $K^*$
distribution amplitudes,
one can estimate the accuracy to which one could extract\cite{Alisugg}
$ |V_{td}/V_{ts}| $
using the ratio of branching rates $\Gamma_{B^+ \rightarrow \rho^+ \, \gamma}/
\Gamma_{B^+ \rightarrow K^{*+} \, \gamma}$.
A conservative estimate  using present bounds\cite{partdata}  for
$V_{bc}V^*_{cd}/V_{bt}V^*_{td}$ and reflecting the sensitivity of this analysis
to the various parameters entering the calculation suggests that  one can
extract
the ratio  $ |V_{td}/V_{ts}| $ to within a factor of two.

\vspace{.5in}
{\centerline{ACKNOWLEDGEMENTS}}
This work was supported in part by the DOE Grant DOE-FG02-93ER-40762.
%\newpage

\newpage
\begin{figure}
\vglue 2.in
\caption{The leading contribution to the decay $B \rightarrow K^* \gamma$.
The wavy line is a photon, the curly line is a gluon.}
\label{diagramlead}
\end{figure}

\begin{figure}
\vglue 2.in
\caption{One of the two ``double--penguin'' contributions to the decay
$B \rightarrow K^* \gamma$.  The second (or photon gluon crossed) graph
is required for gauge--invariance and to obtain a ultraviolet finite result.}
\label{charmloop}
\end{figure}

\begin{figure}
\vglue 2.in
\caption{The leading annihilation graph.  The photon is emitted from the
light quark in the B--meson.}
\label{annihilate}
\end{figure}

\begin{figure}
\vglue 3.5in
\caption{The function $f(r=-2.25)$ (bottom curves) and $f(r=-.5)$ (top curves)
for
the ratio $\Gamma_{B^+ \rightarrow \rho^+ \, \gamma}/
{\Gamma_{B^+ \rightarrow K^{*+} \, \gamma}}$ as a
function of $\Lambda_{QCD}$.   The full lines are the results using the
asymptotic
distribution amplitude for the vector mesons; the dotted lines are those using
the
CZZ distribution.}
\end{figure}

\begin{figure}
\vglue 3.5in
\caption{$f(r=-2.25)$  and $f(r=-.5)$ for $\Gamma_{B^+ \rightarrow
\rho^0 \, \gamma}/
{\Gamma_{B^+ \rightarrow K^{*0} \, \gamma}}$.  Notation same as in Fig. (4).}
\end{figure}
\end{document}